\documentclass[prl,showpacs,superscriptaddress,twocolumn,a4paper]{revtex4}
\usepackage{amsmath}
\usepackage{amsfonts}
\usepackage{amssymb}
\usepackage{graphicx}
\usepackage{color}
\usepackage{array,tabularx}

\def\pj{\phi_J}
\def\<{\langle}
\def\>{\rangle}
\def\Ziso{Z_{\rm iso}}

\def\E0{E_{\rm T=0}}

\begin{document}
\title{High--order jamming crossovers and density anomalies}

\author{Massimo Pica Ciamarra}\email[]{massimo.picaciamarra@spin.cnr.it}
\affiliation{
CNR--SPIN, Dipartimento di Scienze Fisiche,
Universit\`a di Napoli Federico II, I-80126, Napoli, Italy
}
\author{Peter Sollich}
\affiliation{
King’s College London, Department of Mathematics, Strand, London WC2R 2LS, United Kingdom
}

\date{\today}
\begin{abstract}
We demonstrate the existence of high--order jamming crossovers in systems of particles
with repulsive contact interactions, which originate from the collapse of successive
coordination shells.
At zero temperature, these crossovers induce an anomalous behavior of the bulk modulus,
which varies non--monotonically with the density, while at finite temperature they induce density anomalies consisting
in an increased diffusivity upon isothermal compression and in a negative thermal expansion coefficient.
We rationalize the dependence of these crossovers on the softness of the interaction potential,
and relate the jamming crossovers and the anomalous diffusivity through the investigation
of the vibrational spectrum.
\end{abstract}
\maketitle

Repulsive potentials with a finite range are investigated in a variety
of different contexts, including the jamming transition of non thermal
systems, and density anomalies of complex liquids.
In the first case, these potentials are considered as models for the interaction between 
soft particle systems such as emulsions, bubble rafts and granular materials, which 
are not affected by Brownian motion.
These systems undergo a jamming transition marking the onset of mechanical rigidity
when the volume fraction crosses a threshold $\pj$.
See Refs.~\cite{vanHecke2010,Liu2010} for recent reviews. 
At the jamming transition each particle is forced to touch particles
in its first
coordination shell, and the mean contact number per particle jumps to the isostatic
value $\Ziso$, which is the minimum value required for mechanical stability. 
Above $\phi_J$ the excess contact number $\Delta Z = Z-\Ziso$
grows as a power law in $\phi$-$\phi_J$, and is related to a length scale
diverging at the transition as $\Delta Z^{-1}$, known as isostatic length.
The jamming transition is characterized by an abundance of soft vibrational
modes: the density of states $D(\omega)$ satisfies Debye scaling up to a characteristic frequency scaling as $\omega^* \propto \Delta Z$,
above which $D(\omega)$ flattens. 

Similar potentials, and in particular repulsive potentials with a cutoff distance $D$ and with a smaller hard--core 
radius, are also commonly investigated at finite temperature to reproduce the density anomalies
of water and of some other network--forming liquids. See Ref.~\cite{Buldyrev2009} for a recent review. 
A density anomaly is a region
of the phase diagram where the diffusivity increases upon isothermal compression, 
which usually overlaps with a region characterized by a negative thermal expansion coefficient. 
In network--forming liquids, these anomalies are related to soft vibrational modes
characterized by a rigid rotation of tetrahedral structures, known as rigid unit modes~\cite{Dove}.
Even though an abundance of soft modes characterizes both the jamming transition and the density anomalies,
these two phenomena have not been related before, possibly because density anomalies
occur at volume fractions well above $\phi_J$~\cite{Jagla,Stillinger1997,Mausbach,Stanley2005,Berthier2010}.

In this Letter we show that the jamming transition is the first of a series of
high--order jamming crossovers. These occur on increasing the volume fraction as 
particles are forced to make contact with those of subsequent coordination shells.
The geometrical signatures of the jamming crossovers are oscillations in the rate of formation
of new contacts on compression. The mechanical ones include an anomalous volume fraction dependence of
the elasticity of the system, whereby the bulk and shear moduli vary non-monotonically with volume fraction.
We show that density anomalies are the finite temperature counterpart of the jamming crossovers,
and clarify the relation between these phenomena via the study of the soft vibrational modes.

{\it Model --} 
We perform Molecular Dynamics simulations of $50$:$50$ binary mixtures
of particles with diameter $D_l = 1$ and $D_s = D_l/1.4$ and mass $M = 1$.
Two particles $i$ and $j$ interact if they have a positive overlap,
$\delta_{ij} =  D_{ij}-r_{ij} > 0$, where $D_{ij}$ is their average diameter,
and $r_{ij}$ their distance. When this is the case the interaction potential is 
\begin{equation}
 V(r_{ij}) = \frac{1}{\alpha}k \left(\frac{\delta_{ij}}{D_l}\right)^\alpha,
\label{eq:potential}
\end{equation}
where the parameter $\alpha$ controls the softness of the interaction potential,
larger values of $\alpha$ corresponding to softer potentials.
$M$, $D_l$ and $k=1$ are our units of mass, length and energy.
Finite temperature simulations have been performed in the NVT and NPT ensembles, considering systems of $N = 10^3$ particles.
Zero temperature properties have been investigated by quenching random configurations of $N = 10^4$ particles 
to the closest energy minimum via the conjugate--gradient algorithm~\cite{Ohern}.
We have considered a large volume fraction range,
varying from the jamming threshold $\phi \simeq \phi_J$ ($\phi_J \simeq 0.84$ in 2d, $\phi_J \simeq 0.64$ in 3d)
up to $\phi = 3$, and considered the following values of $\alpha$:
$\alpha = 1.25,1.5~\textrm{(Hertzian)},1.75,2~\textrm{(Harmonic)},2.5$ and $3$. The $\alpha = 1$
case corresponds to the Jagla potential with no hard--core repulsion~\cite{Jagla}.
We report in the following results of 2d systems; analogous results hold in 3d.

\paragraph{$T > 0$ properties --} 
\begin{figure}
\includegraphics*[scale=0.35]{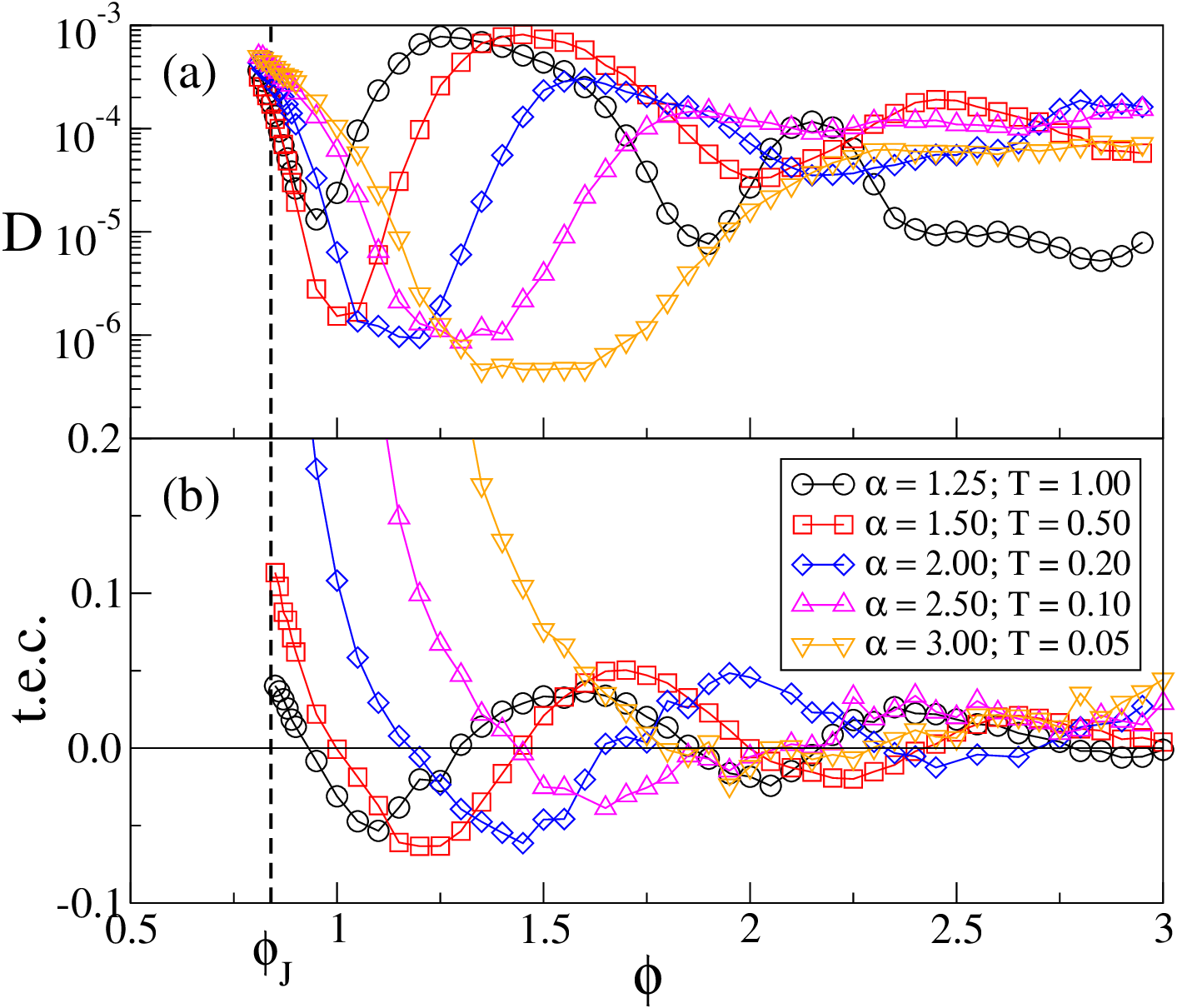}
\caption{\label{fig:diffusion} 
(color online) (a)  Volume fraction dependence of the diffusion coefficient averaged over the two species,
for different interaction potentials. For each potential, the
temperature is chosen so that the minimum 
value of $D$ in the range of $\phi$ studied here is $D \simeq 10^{-6}$.
(b) Thermal expansion coefficient.}
\end{figure}

We start by showing that systems of particles interacting via purely repulsive finite range potentials
are not characterized by a single density anomaly as previously
reported~\cite{Jagla,Stillinger1997,Mausbach,Stanley2005,Berthier2010,Wang2012}, but by a series
of successive density anomalies. Indeed, as illustrated in
Fig.~\ref{fig:diffusion}a, there exist successive
volume fraction ranges in which the diffusivity increases upon isothermal compression.
An analogous result has been observed in models of polymer stars~\cite{foffi2003}.
Likewise, Fig.~\ref{fig:diffusion}b shows 
the existence of multiple volume fraction ranges characterized by a negative thermal expansion coefficient.
The diffusivity anomalies depend on the interaction potential: while
for soft potentials (large $\alpha$)
a single anomaly is observed, more anomalies are
observed on increasing the stiffness of the interaction. Likewise, the potential influences
the number and the strength of the anomalies in the thermal expansion coefficient. 
In the following, we clarify the microscopic origin of these anomalies, and show that
they are the finite temperature counterparts of high--order jamming crossovers.

\paragraph{$T = 0$ properties -- } 
\begin{figure}[t]
\includegraphics*[scale=0.47]{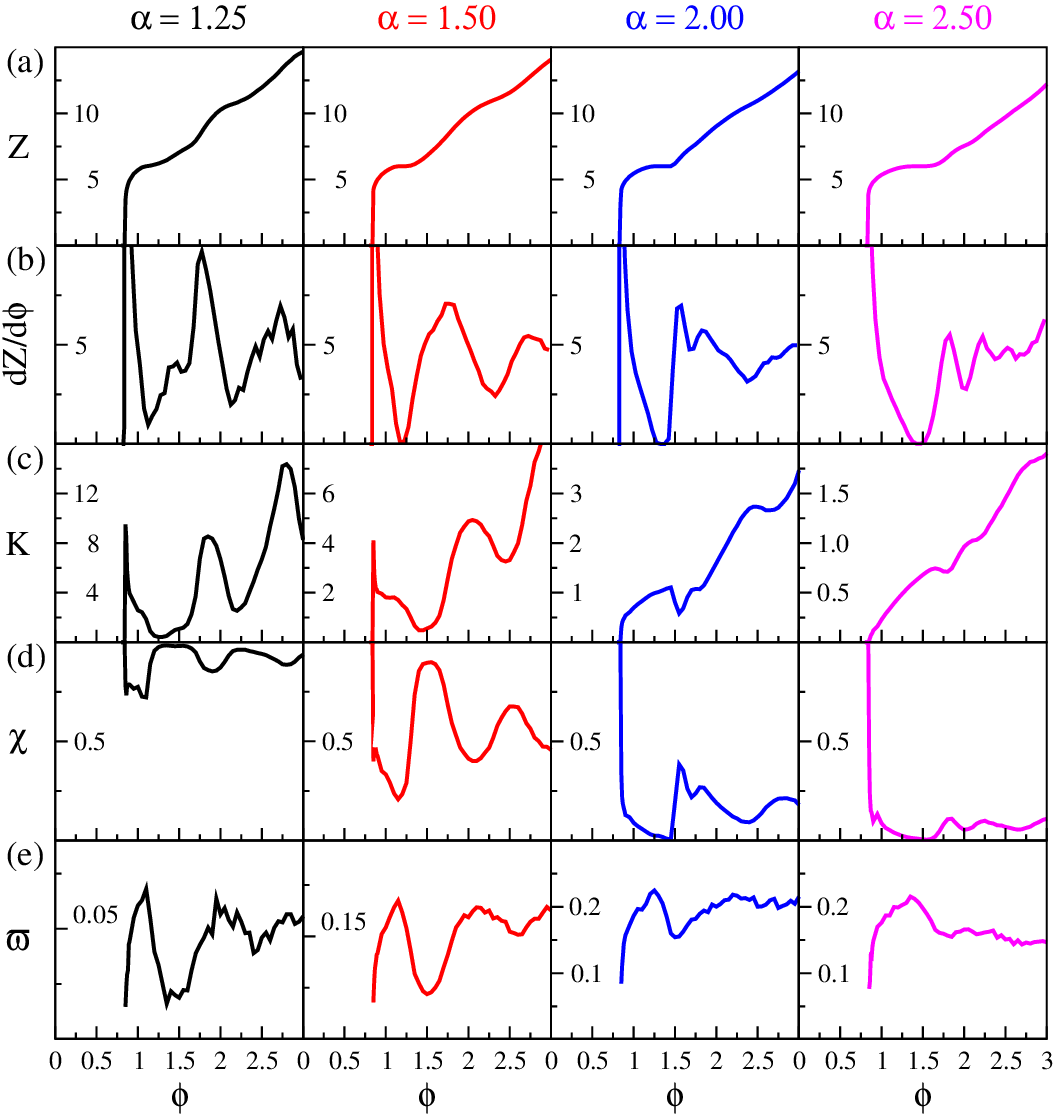}
\caption{\label{fig:compress} (color online) Volume fraction dependence of the mean contact number $Z$ (a), of its volume fraction derivative $dZ/d\phi$ (b),
of the bulk modulus $K$ (c), of the non affinity parameter $\chi$ (d), and of the normalized characteristic frequency $\varpi$ (e). 
Each column refers to a different $\alpha$ value, as indicated. }
\end{figure}
We have investigated the volume fraction dependence of geometrical and mechanical properties
of jammed configurations in a large volume fraction range. To this end we have considered volume
fractions with a spacing of $\Delta \phi = 5\cdot10^{-2}$ far from the
jamming transition and  $\Delta \phi = 10^{-2}$ close to it;
data are averaged
over $50$ independent jammed configurations for each value of $\phi$.
Measurements taken at volume fractions differing by $\Delta \phi$ have been used
to estimate volume fraction derivatives numerically~\cite{note}.
Figs.~\ref{fig:compress}a,b illustrate the volume fraction dependence of the mean contact
number and its first derivative. At the jamming transition, $Z$ jumps to the isostatic value, 
while at higher volume fractions it grows with superimposed oscillations. Consistently,
$dZ/d\phi$ diverges at the jamming transition, and then varies in an oscillating manner. 
As the divergence of $dZ/d\phi$ at the jamming transition corresponds to the collapse of the first coordination
shell, the successive volume fraction ranges in which $dZ/d\phi$ increases can be interpreted as high--order jamming crossovers, 
during which each particle gradually makes contact with neighbors in successive coordination shells. In line with the fact that the
shell structure is lost at high density, the crossovers become smoother as the volume fraction increases, and eventually
$Z$ increases linearly with $\phi$. 
As an aside we note that for soft potentials there is a volume fraction range in which the average contact number is constant, $Z = 6$;
here all particles in contact are V\"oronoi neighbors,
and the value $Z=6$ is fixed by Euler's theorem for planar graphs.
The jamming crossovers also influence mechanical properties such as the bulk modulus $K$,
whose volume fraction dependence is show Fig.~\ref{fig:compress}c. The effect of the crossovers
on $K$ depend on $\alpha$. Indeed, minima and maxima of $K$ and $dZ/d\phi$ are 
closely correlated for small $\alpha$, and become less correlated as $\alpha$ increases, indicating
that the smaller $\alpha$, the greater the influence of the new contacts on the properties of the system.
We finally note that the results of Fig.~\ref{fig:compress} do not support
the conjecture of a transition at a volume fraction $\phi_d > \phi_J$~\cite{Zhao2011}.

\begin{figure}
\includegraphics*[scale=0.35]{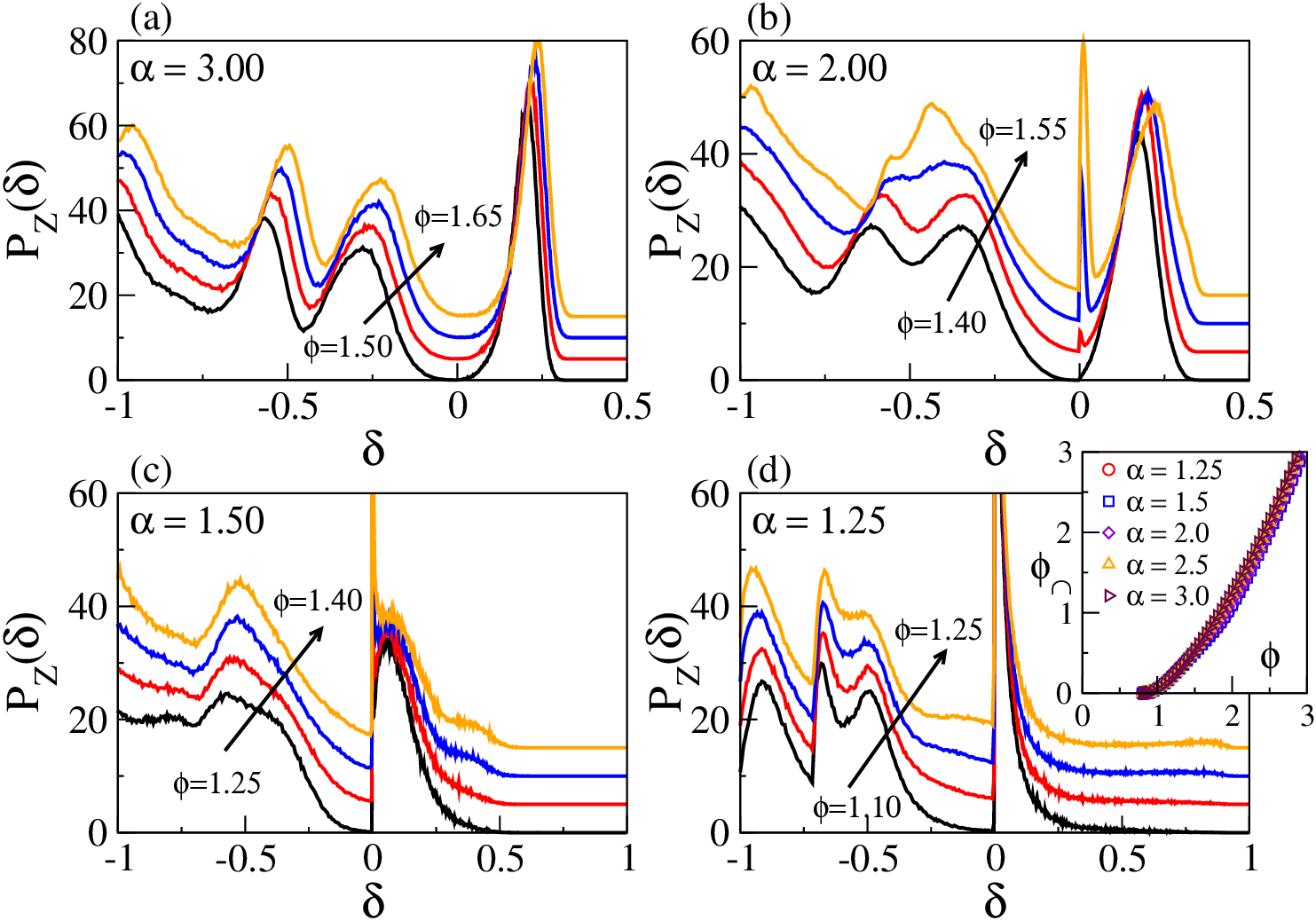}
\caption{\label{fig:fig2} (color online) 
Panels a,b,c: overlap probability distribution function $P_Z(\delta)$ for different values of 
the volume fraction separated by $\Delta \phi = 5 \times 10^{-2}$, as indicated. Data are shifted vertically for clarity.
A negative $\delta$ corresponds to the distance between the surfaces of non--interacting particles.
The inset of panel (d) shows that the intersection volume fraction $\phi_\cap$ is to a good approximation
independent on the softness of the interaction $\alpha$, specified in the legend.
}
\end{figure}

We elucidate the microscopic evolution of the system at the jamming crossovers investigating the volume
fraction dependence of the overlap p.d.f.\ $P_Z(\delta)$, normalized across the interacting particles so that
$\int_0^1P_Z(\delta)d\delta = Z$.
For monodisperse particles $P_Z$ is fully equivalent to the radial distribution function. 
Since $dZ/d\phi$ increases when particles start interacting with those of a new shell, we show in Fig.~\ref{fig:fig2}
$P_Z(\delta)$ at volume fractions immediately following that of the first minimum of $dZ/d\phi$,
for different values of $\alpha$. For $\alpha = 3$ the making of new contacts does not strongly affect the
distribution. $P_Z(\delta)$ is simply shifted towards larger values of $\delta$, which suggests the presence of a mostly affine deformation.
For $\alpha = 2$, on the other hand, the formation of new contacts leads to a transformation of $P_Z(\delta)$, whose net outcome is the creation
of a new peak at small positive $\delta$. This peak shifts towards positive $\delta$ values on further compressing the system.
At smaller $\alpha$, $P_Z(\delta)$ has a diverging peak at $\delta \to 0^+$,
and the transformation of the distribution driven by contact formation leads to an increase of this peak.
The higher order jamming crossovers lead to analogous changes in $P_Z(\delta)$.
These results clarify that the jamming crossovers lead to a rearrangement of the force network
of the system, which is more pronounced the smaller $\alpha$. In particular, we note that 
that on increasing the volume fraction the shell structure observed in $P_Z(\delta)$ for $\delta < 0$,
is transformed into a shell structure for positive overlaps for large $\alpha$,
while conversely it is washed out for smaller $\alpha$. 

To explain the role of the softness of the interaction potential in the shape of
the overlap probability distribution, we introduce the concept of intersection volume fraction $\phi_\cap$.
We define $\phi_\cap$ as the sum of the volumes of intersection of any pair of interacting particles, normalized
by the total volume of the system.
Close to the jamming transition, where the intersection volumes do not overlap,
$\phi = \phi_J + \phi_\cap$, so $\phi_\cap$ is linearly dependent
on $\phi$ and potential independent. In addition, here
$\phi_\cap(\phi)$ scales as the $q$--th moment of $P_Z(\delta)$,
$\phi_{\cap} \propto \langle \delta^q \rangle$,
as the volume of intersection of two particles grows as $\delta^q$ for small $\delta$,
where $q=1.5$ for $d = 2$ ($q=2$ for $d = 3$).
Fig.~\ref{fig:fig2}d (inset) shows that $\phi_\cap$ is also approximately potential independent
at higher volume fractions, so we can regard is as being fixed by $\phi$ throughout. 
The energy of the system, on the other hand, scales as the $\alpha$--th moment of $P_Z(\delta)$,
$E \propto \langle \delta^{\alpha} \rangle$.
Accordingly, constant-$\phi$ energy minimization protocols search for a $P_Z$ distribution
that minimizes $\langle \delta^{\alpha} \rangle$ subject to the constraint of constant $\langle \delta^{q} \rangle$,
or to related constraints away from jamming.
The effect of $\alpha$ on the outcome can be understood by considering the energy $e(\delta_i) = (\delta_i^\alpha + \delta_j^\alpha)/\alpha$
of two overlaps $\delta_i$ and $\delta_j$ related by the constraint of constant intersection volume fraction, 
$\delta_i^q + \delta_j^q = {\rm const}$.
Then for $\alpha > q$ it is energetically 
more favourable for the system to make the overlaps equal, while for $\alpha < q$
the energy is minimized by making them maximally different.
This finding explains the abundance of small contacts in the overlap distribution for small $\alpha$,
and the qualitative change in $P_Z$ as $\alpha$ crosses $q$.

The $\alpha$ dependence of the overlap distribution indicates that, the stiffer the interaction
potential, the more heterogeneous the structure of the system. This suggests an increase of the 
non--affine response of the system on decreasing $\alpha$. We estimate the degree of affinity
of the system by comparing the actual 
bulk modulus $K$ with that computed
in the affine (Born) approximation, $K_{\rm aff}$. If the system responds affinely to the compression,
then $K = K_{\rm aff}$, otherwise $K < K_{\rm aff}$. 
Accordingly, 
the strength of the non--affine response, i.e.\ the relevance of the fluctuation term of the stress tensor~\cite{fluctuation_modulus},
can be quantified via the parameter
\[
 \chi = \frac{K_{\rm aff}-K}{K_{\rm aff}+K}, \qquad 0 \leq \chi \leq 1.
\]
Then $\chi = 0$ when the response is affine, while $\chi \to 1$ when
the response is highly non--affine.
Fig.~\ref{fig:compress}d confirms the expectation that stiffer potentials give rise to a less affine response, 
as the typical value of $\chi$ decreases on increasing $\alpha$.
In addition, the comparison of Figs.~\ref{fig:compress}c and~\ref{fig:compress}d 
reveals the presence of a clear anticorrelation between 
$\chi$ and $K$, which indicates that the jamming crossovers induce an
increase of the degree of non--affinity of the system, much like the jamming transition.
We also note that the oscillations of $\chi(\phi)$ suggest that the non--affine
correlation length that diverges at the jamming transition~\cite{vanHecke2010,Liu2010}
is a non--monotonic function of the volume fraction.

\paragraph{Connecting $T = 0$ and $T > 0$ -- } 
The mechanical anomalies observed at $T = 0$, and the dynamical ones observed at $T > 0$,
have a closely related volume fraction dependence. This suggests that they have the same physical
origin. We elucidate this mechanism by exploiting recent results correlating the thermal~\cite{modes_thermal} and shear
\cite{Tsamados,Manning2011} induced relaxation of glassy systems, and their vibrational modes. 
These studies have shown that the relaxation proceeds through localized events, which occur where
the low frequency modes are quasi-localized. 
In this picture, low--frequency modes are defects allowing for the relaxation of disordered particle systems~\cite{Manning2011}.
In particular, since the activation energy of a soft mode is correlated with its eigenfrequency~\cite{Xu2010},
one expects the dynamics to speed up when the typical frequency of the soft modes decreases.
We have checked this expectation 
by studying the volume fraction dependence of the average eigenfrequency 
$\varpi$ of the lowest modes ($5\%$), normalized
by the average eigenfrequency of all modes. Results are illustrated in Fig.~\ref{fig:compress}e,
for different interaction potentials. 
$\varpi$ exhibits oscillations as the volume 
fraction increases, which are anti--correlated with those of the diffusion coefficient.
In addition, $\varpi$ is also strongly anti--correlated with the parameter $\chi$.
These results confirm the important role of the soft modes in determining
both the relaxation dynamics as well as the mechanical properties of the system.

\paragraph{Conclusions -- } 
We have demonstrated the existence of high--order jamming crossovers in soft
particle systems with increasing density. These occur when particles start to touch neighbors in subsequent coordination
shells. The mechanical manifestation of these crossovers is the anomalous behavior of elastic properties such as the 
bulk and the shear (data not snown) moduli, which vary non-monotonically with the volume fraction. 
Their dynamical manifestation are density anomalies, which include
both an increased diffusivity on compression and a negative thermal expansion coefficient. 
We have related the $T=0$ and the $T>0$ anomalies to the emergence of soft
vibrational modes in correspondence with the crossovers, confirming their important
role in the relaxation dynamics~\cite{modes_thermal,Tsamados,Manning2011}. 
These results suggest that the scaling 
relation between relaxation time and vibrational dynamics found in normal liquids~\cite{Leporini} may also hold for 
anomalous liquids. 
We finally note that the anomalies are observed at the high--order jamming crossovers, and not at the jamming transition,
where the diffusion coefficient decreases monotonically on compression.
One may speculate that this is so because at the jamming crossovers there is a coexistence
of compressed bonds, and of new almost uncompressed bonds. These bonds may play the role of
the two interaction length scales which are known to induce dynamical anomalies in models of water~\cite{Buldyrev2009},
and are responsible for the rigid unit modes leading to a negative thermal expansion coefficient
in network forming systems~\cite{Dove}. These two length scales are not present at the jamming transition.
In this respect, it would be interesting to investigate the spatial structure of the soft modes at the jamming crossover.

\begin{acknowledgments}
MPC thanks the Dept.\ of Mathematics, King's College London for hospitality,
and MIUR-FIRB RBFR081IUK for financial support.
\end{acknowledgments}


\begin{thebibliography}{40}
\bibitem{note}
{
We have checked for some $\alpha$ and $\phi$ that we are working in a region of linear response. 
However, the derivatives we consider are smaller than the ones calculated in the regime of infinitesimal compressions of the system. 
Indeed, quantities calculated in this limit would not capture the presence of avalanche-like rearrangements on compression.
}
\end{thebibliography}
\end{document}